\documentstyle[aps,preprint,epsfig]{revtex}

\begin{document}

\title{Momentum Dependence of Single Particle Potential \\
 in Dirac Brueckner Approach}
\author{C.-H. Lee, T.T.S. Kuo, G.Q. Li and G.E. Brown}
\address{Department of Physics, State University of New
York at Stony Brook, Stony Brook, NY 11794}

\maketitle
 
\begin{abstract}
The momentum dependence of the empirical scalar and vector potentials
needed for describing relativistic heavy-ion collisions at 1 GeV/nucleon 
is compared with that derived from self-consistent Dirac-Brueckner 
calculations using the Bonn-A potential.
Our calculated scalar and vector potentials exhibit 
a weak momentum dependence and their magnitudes decrease
with nucleon momentum, in close similarity with the momentum dependence 
required by the empirical potentials. The effects of explicit momentum 
dependence on the properties of equilibrium nuclear matter are found
to be small.
\end{abstract}
 
 
\newpage

Heavy-ion collisions at various energies open new frontiers and
provide opportunities for nuclear physics. Theoretical description of
heavy-ion collisions is usually  based on transport models.
For heavy-ion collisions at beam energies from a few hundred
MeV/nucleon to a few GeV/nucleon, the reaction dynamics
is mainly governed by the stochastic two-body scattering and
propagation in the mean-field potential. Transport models
that include both effects, such as Boltzmann-Uehling-Uhlenbeck
(BUU) \cite{bert88}, quantum molecular dynamics (QMD) \cite{aich91},
and the relativistic extensions, RBUU \cite{cass90,li94,koli96}
and RQMD \cite{sorge89}, have been extensively used for 
the description of heavy-ion collisions in this energy region.

In the BUU or QMD calculations, the nucleon potential is
usually parameterized in terms of Skyrme-type effective 
interaction \cite{bert88,aich91}. The importance
of the momentum dependence of the nucleon potential
was realized when analysing nucleon flow and pion production
in heavy-ion collisions \cite{gale87,aich87}. 
This has been incorporated into the BUU and QMD approaches
based on various phenomenological parameterizations 
that are constrained by the empirical information 
from proton-nucleus scattering experiments 
\cite{aich91,prak88,gale90}.

On the other hand, the relativistic transport models 
\cite{cass90,li94,koli96} are usually based on the 
Walecka-type effective Lagrangians \cite{qhd}.
In this model, nucleons feel both attractive scalar and repulsive
vector potentials. The corresponding Schr\"odinger equivalent potential, 
that is to be compared with the potentials used in BUU and QMD
calculations, is automatically and explicitly momentum dependent.
In the mean field approximation both the scalar and vector 
potentials are momentum independent. The Schr\"odinger
equivalent potential then increases linearly with nucleon 
kinetic energy, which is in disagreement with what has been
learned from the Dirac phenomenology analysis of proton-nucleus
scattering at intermediate energies \cite{coop93} .

In a recent paper, we have shown that the explicit momentum
dependence in nucleon scalar and vector potential  
is important for the description of nucleon
flow in heavy-ion collisions at 1 GeV/nucleon region
\cite{li97}. In that work, the momentum dependence 
correction was introduced by scaling the momentum 
independent mean field potential with a momentum
dependent function extracted from the Dirac phenomenology. 
There have been similar attempts to reproduce the correct 
momentum dependence of the Schr\"odinger equivalent potential 
by including explicit momentum dependence in the nucleon 
scalar and vector potentials \cite{maru94}.

A main purpose of this paper is to study microscopically the origin
of the momentum dependence in the scalar and vector potentials used 
in relativistic heavy-ion collisions \cite{li97}. 
The self-consitent Dirac Brueckner (DBHF) approach 
\cite{shakin83,mal87,mach89,brock90,li92} is employed. 
It is well-known that the DBHF approach reproduces nicely
the nuclear matter properties starting from realistic nucleon-nucleon
interactions that are constrained by the two-nucleon data.
In most DBHF calculations the scalar and vector potentials
are assumed to be momentum independent. 
This is a good approximation for equilibrium nucleon matter,
where the nucleon momenta involved are not very large.
The momentum dependence was studied by ter Haar and Malfliet
\cite{mal87}, and they found a relatively weak 
momentum dependence for both the real and imaginary parts
of the scalar and vector potentials. 
Recently, Sehn {\it et al.} \cite{sehn97} studied
the momentum dependence of the nucleon scalar and vector potentials
in the first-order approximation. They first carried
out the momentum independent DBHF calculation as in Ref.
\cite{mach89}. The scalar and vector potential so obtained
were identified with a nucleon at Fermi surface, namely $|{\bf k}|
=k_F$. The potentials for momenta below and above $k_F$ were
then calculated using the formalism of Horowitz and Serot 
\cite{horo87}. The self-consistency in the DBHF calculation
has thus been neglected in Ref. \cite{sehn97}. We will compare
our results to those of Refs. \cite{mal87,sehn97}.

In this work, we follow essentially the DBHF formalism of
Ref. \cite{mach89}. The basic quantity is the $\tilde G$
matrix that satisfies the in-medium Thompson equation,
\begin{eqnarray}
\tilde G (q^\prime,q | P,\tilde z) &=&
\tilde V (q^\prime,q )
+\int \frac{d^3k}{(2\pi)^3} \tilde V (q^\prime,k )
  \left(\frac{\tilde m (k)}{\tilde E(k)}\right)^2 
  \frac{Q(k,P)}{2\tilde E(q) -2 \tilde E (k) }
\tilde G (k,q | P,\tilde z),
\end{eqnarray}
where $\tilde E (k)\equiv \sqrt{{\tilde m}^2+(P/2+k)^2}$, $P$ is c.m.
momentum, $\tilde z = 2 \tilde E(q)$ and $\tilde m (k)=m+U_S(k)$.
The single-particle potential can be calculated once the 
$\tilde G$ matrix is known,
\begin{eqnarray}
\Sigma_{DB}(k) &=& Re \int_0^{k_F} d^3q  
  \left(\frac{\tilde m(q)}{\tilde E(q)}\right)
  \left(\frac{\tilde m(k)}{\tilde E(k)}\right)
 \left\langle k q \left| \tilde G (\tilde z) \right| kq-qk\right\rangle .
\label{equdb}
\end{eqnarray}
This potential is usually parameterized in terms of scalar and
vector potential,
\begin{eqnarray}
\Sigma (k) & \approx & \frac{\tilde m(k)}{\tilde E (k)} U_S (k) +U_V (k).
\end{eqnarray}
In Ref. \cite{mach89}, both the scalar and vector potentials
are assumed to be momentum independent. Thus for a given density,
these potentials can be determined by choosing two momenta
in Eq. (3), usually $0.7k_F$ and $k_F$. In the momentum
dependent case as we consider here, we have two unknowns 
($U_S$ and $U_V$) for each given momentum, which cannot be
uniquely determined without introducing further approximation.
The customary way to do this is to assume some functional
dependence on momentum for these potentials. For example, 
in Ref. \cite{muth87}, it was assumed that $U_{S,V} \approx
A_{S,V}+B_{S,V}k^2$. Since the real part of the single particle 
potential vanishes at very high momentum,
we expect the following parameterization to be 
more appropriate
\begin{eqnarray}
U_{S,V} (k) &=& \frac{(U_{S,V})_0}{1+\alpha_{S,V} (k/k_F)^{\beta_{S,V}}}.
\label{usvpar}\end{eqnarray}
Actually, in Ref. \cite{li97}, this kind of functional dependence
was found to fit the Dirac phenomenology rather well.
One of the purposes of this work is thus to
study whether this kind of simple parameterization
is supported by microscopic DBHF calculation.
Numerically we find that it is rather difficult to
determine the six parameters altogether in the
minimization procedure. We have checked the variation
of $\beta_{S,V}$ around 0.5, and find that 
$\beta_{S,V}\approx 0.5$ provides a smooth self-consistent 
procedure, and the final results are not very sensitive to
some small change in $\beta_{S,V}$.
The remaining four parameters are then determined by 
a minimization procedure,
\begin{eqnarray}
\chi^2 &=& \int_0^{k_F} d^3 k \left(\Sigma_{DB}(k)-\Sigma (k)\right)^2.
\label{eqchisq}\end{eqnarray}

\begin{figure}
\centerline{\epsfig{file=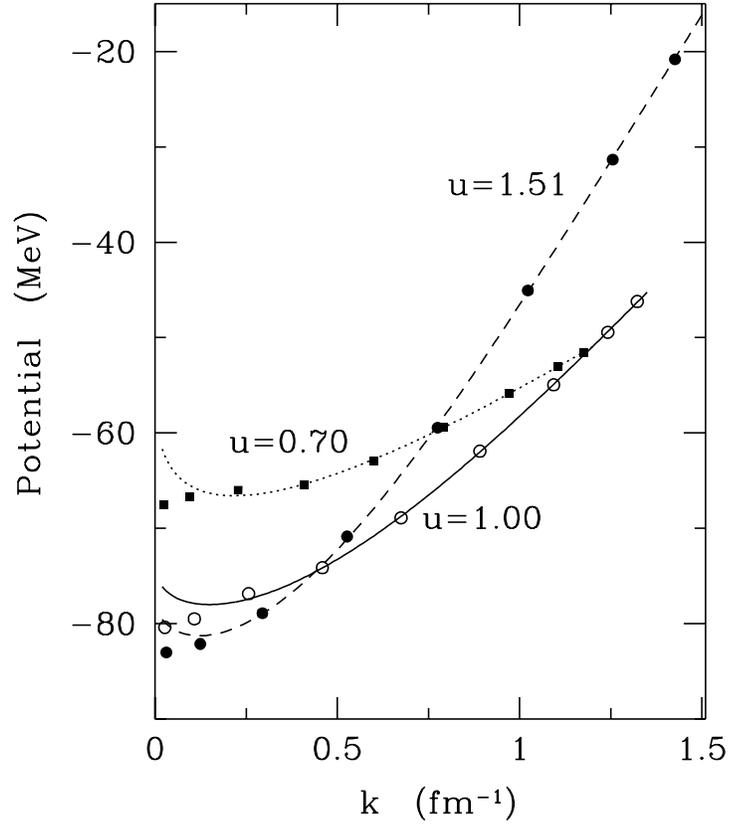,height=12cm}}
\caption{Momentum dependence of single particle potential.
The solid symbols are the results of DBHF calculation and the
lines are from Eqs. (3) and (4).}
\label{figmsp}
\end{figure}

Our self-consistent procedure is the following. For a given density
as specified by Fermi momentum $k_F$, we choose reasonable initial
values for $(U_{S,V})_0$ and $\alpha _{S,V}$. These, together with
Eq. (4) provide $U_{S,V}$ for all momenta. These potentials are then
used as inputs in Eq. (1) to determine the $\tilde G$ matrix,
from which we obtain the single-particle potential $\Sigma _{DB}$
using Eq. (2). Our new set of $(U_{S,V})_0$ and $\alpha _{S,V}$
can then be determined by minimizing Eq. (\ref{eqchisq}) 
against $\Sigma _{DB}$. This procedure is continued until desired 
self-consistency is achieved. The quality of our self-consistency 
procedure can be seen in Fig. \ref{figmsp} where we compare 
the single-particle potential from the DBHF calculation 
(solid symbols) and from  Eqs. (3) and (4). There are slight 
deviations near $k=0$ because in our definition of $\chi^2$, 
Eq. (5), the contributions from small momenta are suppressed.
The self consistent parameters are summarized in Tab. \ref{tab1}.

\begin{figure}
\centerline{ \epsfig{file=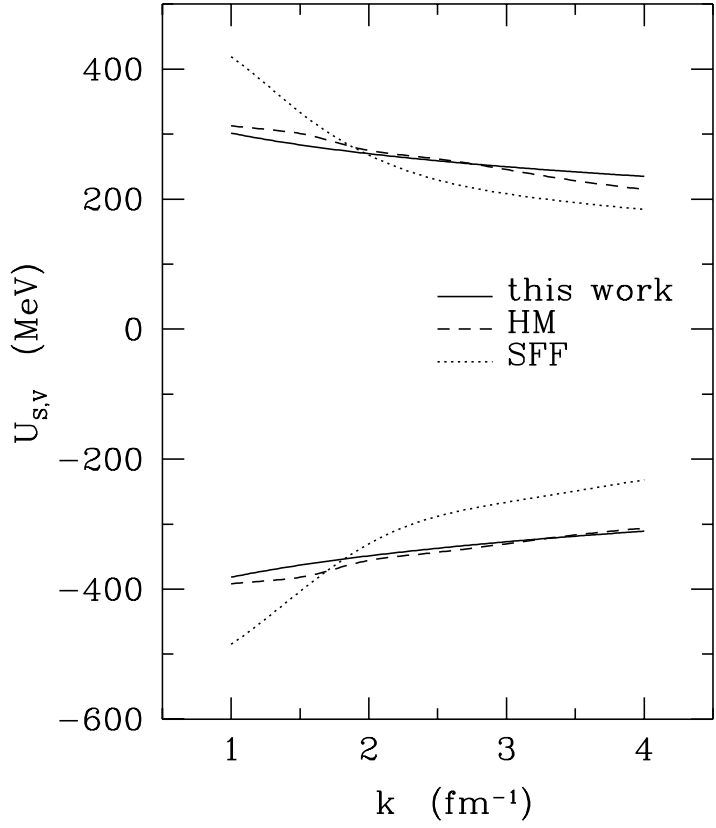,height=12cm}}
\caption{nucleon scalar and vector potential as a function of
nucleon momentum at normal nuclear matter density.
The solid curves are our results, while the dashed and
dotted curve are the results from Refs. \protect\cite{mal87} (HM)
and \protect\cite{sehn97} (SFF), respectively.}
\label{figusv}
\end{figure}

\begin{figure}
\centerline{\epsfig{file=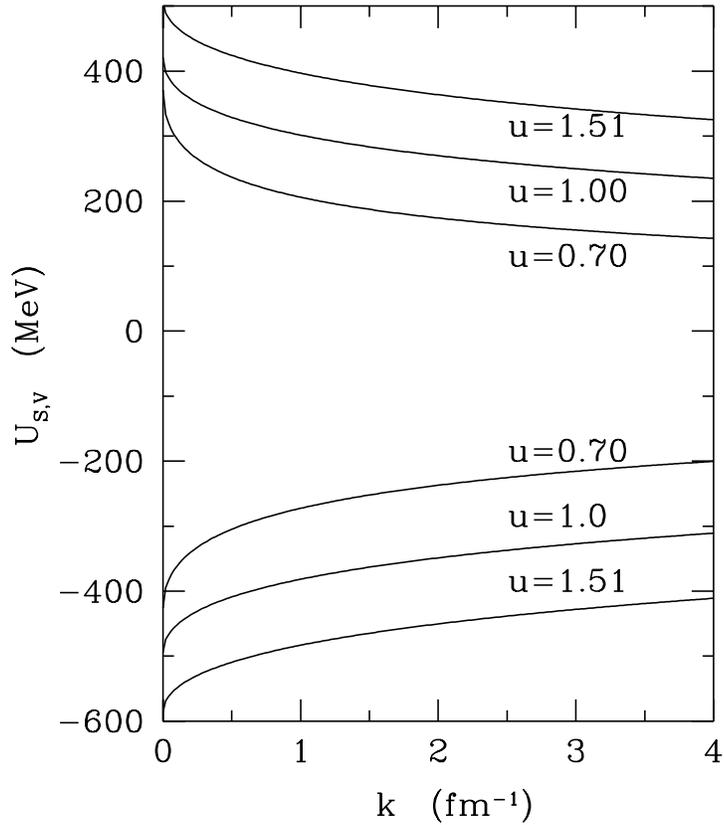,height=12cm}}
\caption{nucleon scalar and vector potential as a function of
nucleon momentum at several different densities.}
\label{figusv2}
\end{figure}

In Fig. \ref{figusv} we show our results for the momentum
dependence of the scalar and vector potentials at normal nuclear
matter density. Both are seen to decrease in magnitude as 
nucleon momentum increases. Also shown in the figure are the 
results of ter Haar and Malfliet (HM) \cite{mal87}
and those of Sehn {\it et al.} (SFF) \cite{sehn97}.
Our results are very close to those of ter Haar and Malfliet, 
which supports our parametrization functional,
Eq. (\ref{usvpar}). The results of Sehn {\it et al.} 
show a much stronger momentum dependence than ours and those of
ter Haar and Malfliet \cite{mal87}, because
of the lack of the self-consistency. 
We show in Fig. \ref{figusv2} the momentum dependence
of the nucleon scalar and vector potentials at several
different densities.

\begin{figure}
\centerline{\epsfig{file=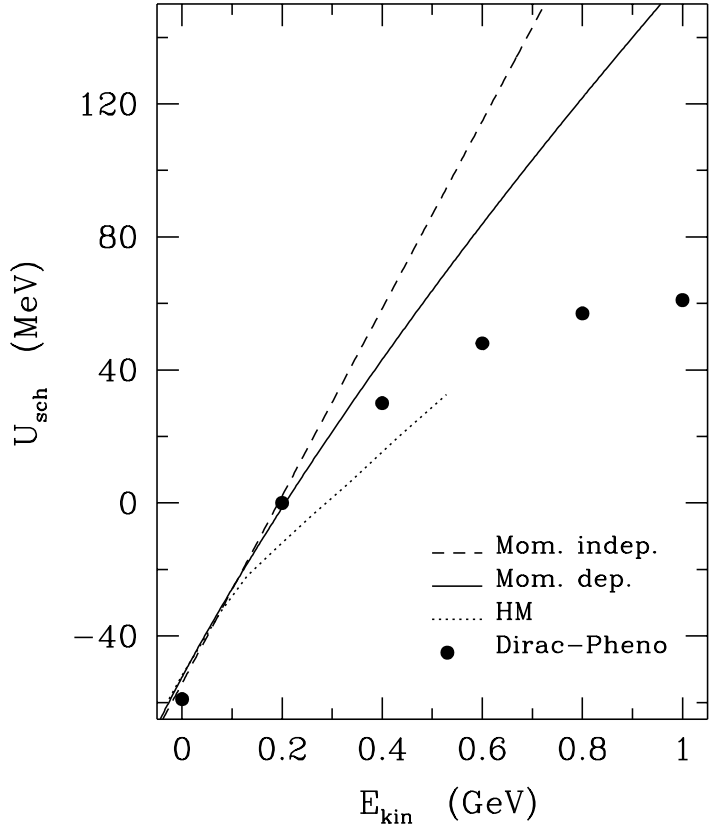,height=12cm}}
\caption{Schr\"odinger equivalent potential in the DBHF calculation
with and without explicit momentum dependence. The solid circles are
extracted from the Dirac phenomenology analysis \protect\cite{coop93}.
The results of Haar and Malfliet (HM) \protect\cite{mal87} 
are plotted in dotted line.}
\label{figsch}
\end{figure}

{}From the scalar and vector potential we can calculated
the Schr\"odinger equivalent potential given by
\begin{eqnarray}
U_{sch}(k) = U_S (k) + \left(\frac{E_{kin}(k)}{m}+1\right) U_V(k)
  + \frac{U_S(k)^2 -U_V(k)^2}{2 m} ,
\end{eqnarray}
where $E_{kin}$ is the nucleon kinetic energy.
The results are shown in Fig. \ref{figsch}. The solid symbols
are the Schr\"odinger equivalent potential extracted from
the Dirac phenomenology analysis \cite{coop93}.
The solid and dashed curves are obtained with and without
explicit momentum dependence in the scalar and vector
potentials. The dotted is from ref. \cite{mal87}.
It is seen that the explicit momentum dependence in the
scalar and vector potentials improves, to some extent,
the agreement with the Dirac phenomenology. 
The Schr\"odinger equivalent potential from our
self-consistent calculation still show stronger 
kinetic energy dependence than that of the Dirac
phenomenology. The onset of inelastic processes
at high beam energies might explain part of this discrepancy.
 
\begin{figure}
\centerline{\epsfig{file=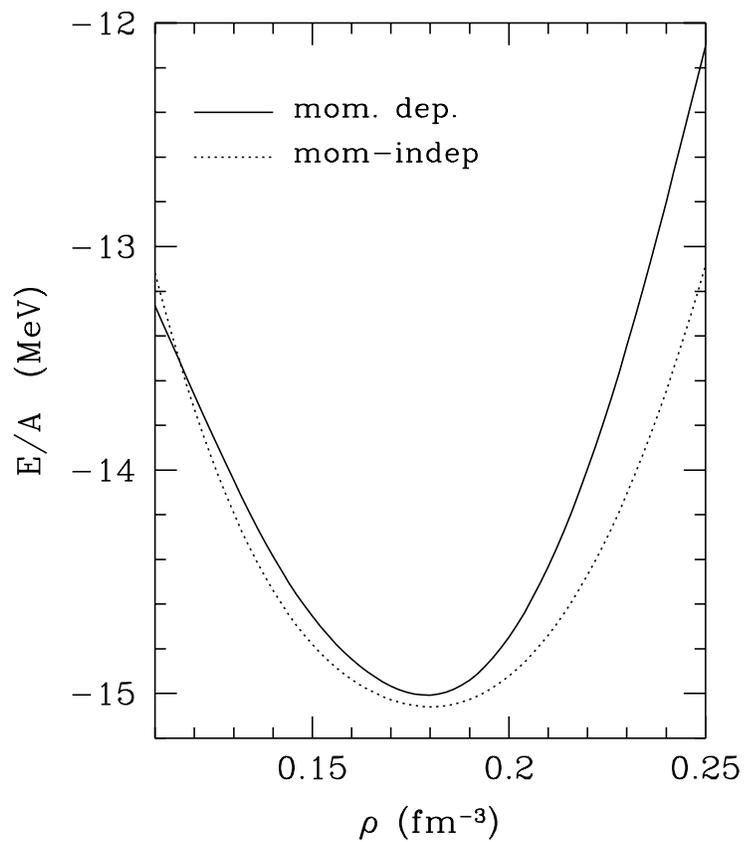,height=12cm}}
\caption{Binding energies in the DBHF calculation
with and without explicit momentum dependence.}
\label{figbe}
\end{figure}

It is also of interest to see how the nuclear equation of state
are affected by the explicit momentum dependence.
\begin{eqnarray}
E/A &=&\frac{1}{\int_0^{k_F} d^3k} \int_0^{k_F} d^3k
\left( \frac{M \tilde M(k) +k^2}{\tilde E(k)} + \frac 12 \Sigma (k) \right)
- M .
\end{eqnarray}
In Fig.\ref{figbe}, the binding energies are plotted as a function of
density. The solid and dashed curves are obtained with and
without explicit momentum dependence in the scalar and
vector potentials. It is seen that the effects of the
explicit momentum dependence on equilibriu nuclear matter
is insignificant. The saturation density in both cases are about
$\sim 0.18 {\; {\rm fm}}^{-3}$. The binding energy 
decreases by about 0.2 MeV when explicit momentum dependence
is included.  

In summary, we studied the momentum dependence of nuclear scalar 
and vector potentials self-consistently in the formalism of Dirac 
Brueckner approach. The magnitudes of both potentials have been
found to decrease slowly with increasing nucleon momentum or
kinetic energy. This improves, to some extent, the agreement 
with the Dirac phenomenology. The effects of explicit momentum 
dependence on the properties of equilibrium nuclear matter were 
found to be small.

\section*{Acknowledgement}
We would like to thank M. Prakash, M. Rho and C.M. Ko for 
helpful discussions. We also appreciate R. Machleidt for sending us 
the code of the relativistic Bonn potential and helpful discussions.
This work was supported by the U.S. Department of Energy under
grant no. DE-FG02-88ER40388. The work of CHL was supported in 
part by Korea Science and Engineering Foundation.

\newpage
\begin{table}
$$
\begin{array}{|c||c|c|c|c|}
\hline
u (\rho/\rho_0) & U_{s0} (MeV) & \alpha_s & U_{v0} (MeV) & \alpha_v \\
\hline
 0.70 & -426.2505 & 0.6178 & 370.6928 & 0.8754 \\
 0.85 & -474.3686 & 0.4283 & 408.2838 & 0.5805 \\
 1.00 & -494.2272 & 0.3426 & 420.5226 & 0.4585 \\ 
 1.12 & -499.1539 & 0.2790 & 420.0603 & 0.3687 \\ 
 1.24 & -518.8160 & 0.2566 & 436.9299 & 0.3361 \\
 1.37 & -537.0683 & 0.2357 & 453.5746 & 0.3050 \\
 1.51 & -586.0774 & 0.2652 & 508.8679 & 0.3515 \\
\hline
\end{array}
$$
\caption{Self-consistent parameters with $\rho_0=0.166{\; \rm fm}^{-3}$.}
\label{tab1}
\end{table}

\newpage

\newpage

\newpage

\newpage

\newpage

\end{document}